\documentclass{PoS}
\usepackage{graphicx}
\usepackage{url}

\title{VLBI detection of internal shocks in nova V959 Mon}

\ShortTitle{VLBI detection of internal shocks in nova V959 Mon}

\author{\speaker{Jun Yang}$^{1,2,3}$, Zsolt Paragi$^2$, Tim, J. O'Brien$^4$, Laura Chomiuk$^5$, Justin D. Linford$^5$%
        \thanks{The EVN is a joint facility of European, Chinese, South African and other radio astronomy institutes funded by their respective national research councils. The EVN and e-VLBI research infrastructures were supported by the European Commission Seventh Framework Programme (FP/2007-2013) under grant agreements nos 283393 (RadioNet3) and RI-261525 (NEXPReS).  e-MERLIN is operated by The University of Manchester at Jodrell Bank Observatory on behalf of the Science and Technology Facilities Council. } \\
        
       $^1$Department of Earth and Space Sciences, Chalmers University of Technology, Onsala Space Observatory, SE-439 92 Onsala, Sweden. 
       E-mail: \email{jun.yang@chalmers.se} \\
       $^2$Joint Institute for VLBI in Europe, Postbus 2, NL-7990 AA Dwingeloo, The Netherlands \\
       $^3$Shanghai Astronomical Observatory, Chinese Academy of Sciences, 80 Nandan Road, 200030 Shanghai, P.R. China \\
       $^4$Jodrell Bank Centre for Astrophysics, Alan Turing Building, University of Manchester, Manchester M139PL, UK \\
       $^5$Department of Physics and Astronomy, Michigan State University, East Lansing, Michigan 48824, USA \\
}  

\abstract{
V959~Mon is a classical nova detected at GeV $\gamma$-ray wavelengths on 2012 June 19. While classical novae are now routinely detected
in gamma-rays, the origin of the shocks that produce relativistic particles has remained unknown. We carried out electronic European VLBI Network (e-EVN) 
observations that revealed a pair of compact synchrotron emission features in V959~Mon on 2012 Sep 18. Since synchrotron emission requires strong shocks as well, we identify these features as the location where the gamma rays were produced. We also detected the extended ejecta in the follow-up EVN observations. They expanded much faster in East-West direction than the compact knots detected in the aforementioned e-EVN measurements. By comparing the VLBI results with lower resolution images obtained using e-MERLIN and the VLA -- as reported by Chomiuk et al. \cite{chomiuk14} -- it appears that 1) influenced by the binary orbit, the nova ejecta was highly asymmetric with a dense and slow outflow in the equatorial plane and low-density and faster ejecta along the poles; and 2) the VLBI knots were related to shocks formed in the interaction region of these outflows.}

\FullConference{12th European VLBI Network Symposium and Users Meeting\\
		 7-10 October 2014\\
		 Cagliari, Italy}

\begin{document}

\section{Introduction}
\label{sec1}

A nova is a thermonuclear explosion on the surface of a white dwarf star in a binary system. The outbursts of novae are multi-wavelength phenomena on time scale of days to years. Their luminosity rises to the maximum first in optical, then in UV, and finally in the X-rays as their photospheric effective temperature increases. A rise and decay has been observed in the infrared in some novae due to the formation of optically thick dust at the later stage. In the radio thermal bremsstrahlung/free-free emission comes from the ionised gas. The radio light is initially optically thick because of free-free absorption. Multi-frequency observations of the absorbed radio spectrum will enable the most reliable estimate of ejected mass \cite{seaquist12}. 

Some novae also have significant non-thermal radiation in radio and Gamma-ray spectral domains.  Since the first radio detection of RS Oph in outburst \cite{padin85}, a non-thermal origin due to the interaction with the external, pre-existing wind driven by the giant companion has also been included in the explanation of radio emission in the sub-class of novae with giant companions. Later, very long baseline interferometry (VLBI) observations have directly revealed an asymmetric shock wave in the 2006 outburst of the recurrent nova RS Ophiuchi \cite{obrien06}.  The first nova detected in $\gamma$-rays was the recurrent nova V407 Cygni \cite{abdo10}. The formation of the $\gamma$-ray emission is again related to the shocks with the high-density winds. According to this scenario, no GeV $\gamma$-ray emission is expected in classical novae that have main sequence stars rather than red giants as their companions. Surprisingly, $\gamma$-ray emission has been recently found in a group of classical novae including nova V959 Mon \cite{fermi14}. Similarly to synchrotron radio emission, $\gamma$-rays require a population of relativistic particles that are naturally produced in strong shocks. This provides a way to identify the likely locations of gamma-ray producing shocks, by detecting compact synchrotron emitting regions with VLBI. 

Nova V959 Mon was discovered by Fermi-LAT \cite{fermi14} in gamma-rays on 2012 Jun 19 (hereafter reference as day 0). The detectable $\gamma$-ray activity lasted about 21 days. Its companion is most likely a main-sequence, early-type K star \cite{munari13}.  Karl G. Jansky Very Large Array (VLA) observations of V959 Mon on days 12 and 16 revealed a flat spectrum at frequencies $\leq6$~GHz, supporting the existence of the synchrotron radiation \cite{chomiuk14}. To further locate these synchrotron radiating regions, we carried out target of opportunity observations with the European VLBI network (EVN). In this paper, we present the EVN imaging results. 
   
\section{EVN observations and data reduction}
\label{sec2}

We performed EVN observations of nova V959~Mon at 5.0 GHz at five epochs (project codes: RO005, RO006, EO011) from 2012 Sep 18 to 2013 Jan 15, for about 6--8 hours per epoch. For phase-referencing we selected J0645+0541 as calibrator. The used nodding cycle time was $\sim4$ minutes (calibrator: $\sim1$ min, target: $\sim3$ min). We also inserted some short scans on a weak flat-spectrum source J0639+0601, 11 arcmin apart,  in order to verify the phase-referencing reliability across these epochs. DA193 was observed for a few scans to calibrate out the instrumental bandpass response. The e-transferred data were correlated in real time at the Joint Institute for VLBI in Europe (JIVE). 

The data processing was done in AIPS, while the final images were made in Difmap. With respect to J0645+0541, we measured the mean J2000 position of the secondary calibrator  $\mathrm{RA}=06^\mathrm{h}39^\mathrm{m}06.\!\!^\mathrm{s}84006$,  $\mathrm{Dec}=+06^\circ01^\prime33.\!\!^{\prime\prime}6864$ in the first four epochs.  The statistical position error is $1\sigma=0.43$ mas in RA and $1\sigma=0.70$ mas in Dec. Note that the initial position of J0639+0601 had a quite large offset ($-2.\!\!^{\prime\prime}613$) in Dec in the first epoch and the large position error was corrected in the later observing schedules. J0639+0601 is a point-like source with a total flux density $36\pm2$mJy and a correlation amplitude $\sim25$~mJy on the long baselines. We also ran phase self-calibration on the secondary calibrator and transferred its phase solution to the data of V959~Mon. Note that the same input source model is used during the phase self-calibration to minimise the position error in the secondary reference source itself.   

\begin{figure*}
  \centering
  \includegraphics[width=0.99\textwidth, clip]{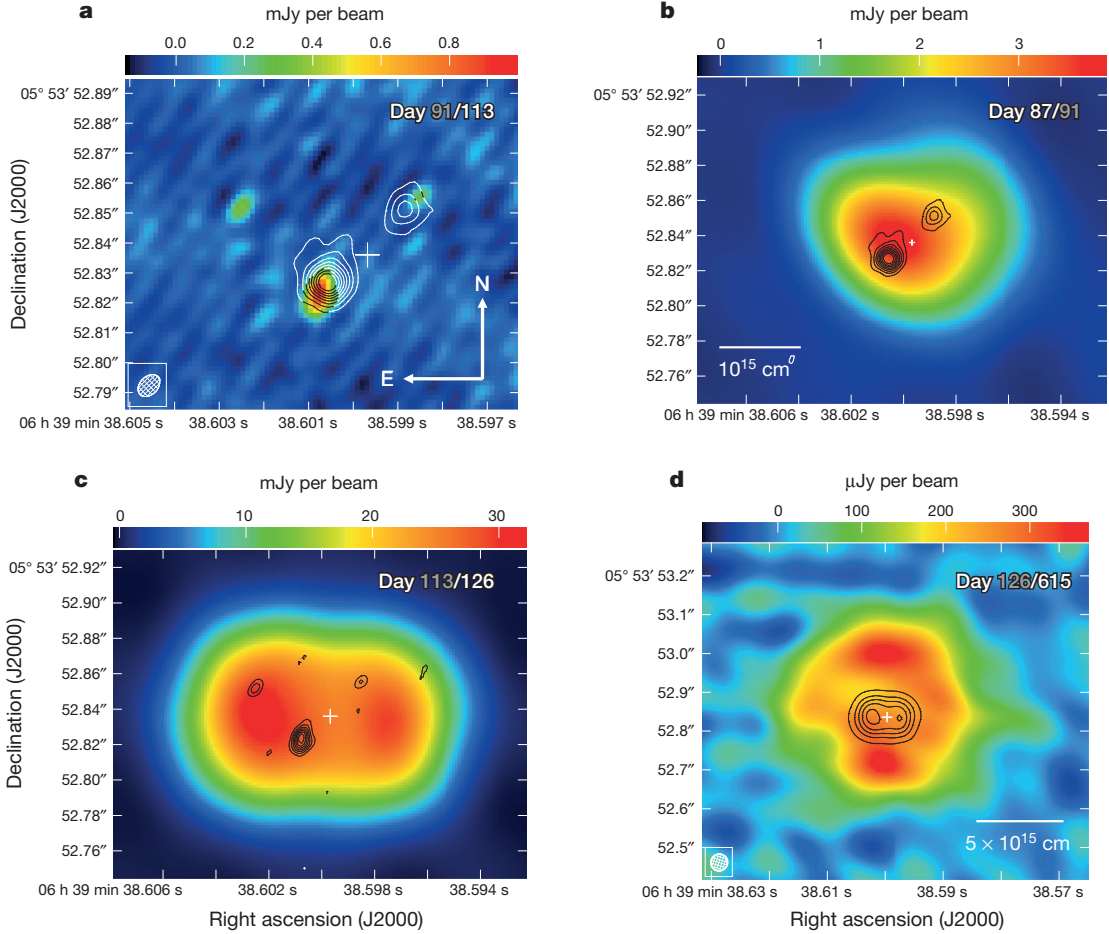}  \\
  \caption{The multi-resolution radio images of nova V959 Mon \cite{chomiuk14}. (a), EVN detection of radio knots compact on mas scale and expanding from 91 (contours) to 113 days (pseudo colour). (b), EVN contour image of day 91 over-plotted on the e-MERLIN colour image of day 87, (c), EVN contour image of day 113 over-plotted on the VLA 36.5GHz colour image of day 126. (d), the VLA images of day 126 at 36.5 GHz vs day 615 at 17.5 GHz. The VLA and e-MERLIN revealed the large scale thermal emission region, which displayed a 90 degree flip of radio structure due to the non-spherical distribution of ejecta material.  } \label{fig1}
\end{figure*}

\section{VLBI detection of compact knots in V959 Mon}
\label{sec3}
The EVN imaging results are shown in Figure~\ref{fig1}a. There are two radio knots detected in the first epoch and three knots in the second epoch. According to their relative position, we named them as components NW, SE, and E.  All these knots have a good signal to noise ratio $>$~10.  The image sensitivity is about $0.03$~mJy/beam in the two maps. The synthesised beam is $9.2\times5.5$~mas at position angle $-38.\!\!^\circ0$ in the first epoch and $8.4\times6.0$~mas at position angle $-44.\!\!^\circ3$ in the second epoch. 

We fitted a Gaussian model to the uv-data for each compact radio knots. Their positions and flux densities have been reported in the paper of Chomiuk et al. \cite{chomiuk14}. The EVN observations show that components NW and SE were expanding away from each other at a separation speed of $0.50\pm0.03$ mas~d$^{-1}$ in NW--SE direction.  All these radio knots are compact on the mas scale. We also attempted to fit each radio knot with a model consisting of a few point sources in order to search for compact substructures. Only component SE can be well modelled by two point sources (day 91: $0.74\pm0.07$ and $0.35\pm0.04$~mJy, day 113: $0.78\pm0.08$ and $0.58\pm0.06$ mJy). The two point sources had a separation varying from $3.4\pm0.7$~mas to $5.7\pm0.7$~mas in NS direction. Component SE likely represents a linear emission region. The VLBA follow-up observations \cite{chomiuk14} also detected these compact radio knots at both 1.6 and 5 GHz, in agreement with the EVN imaging results.  

The VLA and e-MERLIN imaging results are also shown in Figure~\ref{fig1}b, c, and d. Thanks to their large beam,  the emission resolved out by VLBI is fully recovered. The radiation intensity distribution had evolved into a clearly double-lobed morphology in the VLA images on day 126. The VLBI knots lay roughly halfway between the directions of the early and later expanding axes. Comparing with the restored flux density by the WSRT observations, the VLBI knots comprised 20\% of the flux density on day 91 and 11\% on day 113.  There was no compact radio emission detected at the last three epochs because of the low peak brightness and significant expansion. The upper limit of peak brightness is $5\sigma\sim$0.25~mJy/beam on 14 Nov 2012, $5\sigma\sim0.15$~mJy/beam on 4 Dec 2012, and $5\sigma\sim0.18$~mJy/beam on 15 Jan 2013. 

\begin{figure*}
  \centering
  \includegraphics[origin=br, width=0.325\textwidth, clip]{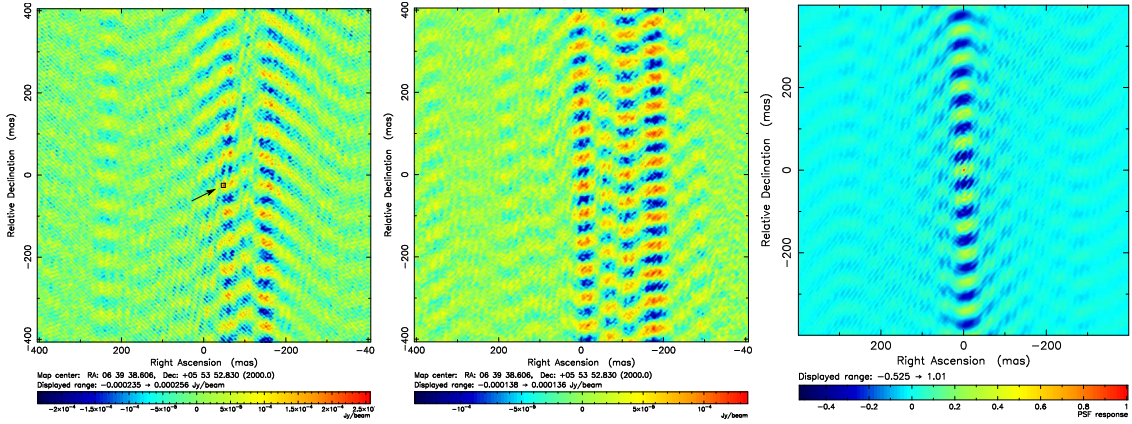}  
  \includegraphics[origin=br, width=0.325\textwidth, clip]{nova-mon_dec04.ps} 
  \includegraphics[origin=br, angle=-90, width=0.325\textwidth, clip]{beam.ps}     \\
  \caption{The naturally weighted dirty maps produced in the EVN observations of nova V959 Mon on 2012 Nov 14 (left) and 2012 Dec 4 (middle) and the synthesised dirty beam (right).   The two stripes in North-South direction indicates the existence of two extended radio features.  The small black square marked by a black arrow in the left panel shows the map peak-brightness position, i.e. the possible position of the Eastern extended radio feature. }\label{fig2}
\end{figure*}

\section{The nature of VLBI-detected knots -- internal shocks} 
\label{sec4}
The compact knots in V959 Mon are due to non-thermal synchrotron radiation rather than thermal free-free radiation, and they most likely trace the shocks responsible for the$\gamma$-ray emission. According to our VLBI imaging results, these knots had a peak brightness temperature $\sim2\times10^6$ K, clearly higher than the typical brightness temperature $\leq10^5$~K observed in a free-free emission region. Furthermore, the VLA observations revealed \cite{chomiuk14} a flat spectrum between 1.3 and 6 GHz on day 16 and a rising spectrum between 1.3 and  36 GHz on day 145. 

There was also a hint for the existence of diffuse radio emission region between the compact knots in the residual intensity map at the first two epochs. The compact knots were not detected at the third and fourth epochs, while these diffuse regions became slightly brighter. The naturally weighted dirty maps are shown in Figure~\ref{fig2}. A proper imaging of the data was not possible due to the lack of short $uv$-spacings. The peak brightness is 0.26 mJy/beam in the left map and 0.14 mJy/beam in the middle map. The dirty beam is characteristic of a vertical stripe formed by the red/positive and blue/negative side lobes. Through searching for a regular pattern similar to the dirty beam, we have identified two extended radio features showing as two vertical stripes in the third epoch at $\Delta\mathrm{RA}=-57\pm2$ mas and $\Delta\mathrm{RA}=-158\pm2$ mas,  and in the fourth epoch at $\Delta\mathrm{RA}=-11\pm2$ mas and $\Delta\mathrm{RA}=-175\pm6$ mas. Their position in Dec. can not be reliably measured due to the strong side lobes of the synthesised beam, but the continuing expansion is clearly visible. We estimated an expansion rate of $3.1\pm0.4$~mas~d$^{-1}$ in EW direction. The large uncertainty is caused by their large size (12~--~30 mas). The VLA observations gave an average expansion rate of $0.64\pm0.04$ mas~d$^{-1}$ at 13.5 GHz \cite{chomiuk14}, almost a factor of five lower than that observed by the EVN. The two measurements do not conflict with each other considering the large difference in the resolution, the restored flux density, and the observing frequency. Actually, their co-existence significantly strengthens the scenario that the VLBI knots/shocks are formed by the differential velocity inside the nova shell. 

The apparent rotation of the elongation axis of the radio morphology between day 126 and 615 in Figure~\ref{fig1}d was also caused by the nova shell expanding faster and getting optically thin earlier along EW direction.  If a source is optically thin at the observing frequencies, the measured expansion rate is decreasing with the increase of the observing frequency.  As expected, the frequency dependence is observed only in EW direction by the VLA observations during days 133~--~196 \cite{chomiuk14}.

The observed structure at the various angular scales were interpreted by Chomiuk et al. \cite{chomiuk14} as follows. V959~Mon entered early in the nova outburst when the expanding nova shell reached the secondary star. The binary system transferred some of its angular momentum to the surrounding material via viscous interaction, resulting in a denser outflow in the equatorial plane.  Meanwhile, the white dwarf powered a fast wind that interacted with the nova shell and was funnelled toward the low-density polar directions. At the interaction regions shocks were formed due to the different outflow velocities. In the shocks relativistic particles were produced that were responsible for the observed gamma rays. These regions appeared later as synchrotron self-absorbed radio knots.

\end{document}